# Towards Security Enhancement of Blockchain-based Supply Chain Management

L. Al-Alawi, R. Al-Busaidi, A. K. Shaikh

Department of Information systems, Sultan Qaboos University, Muscat, Oman,
s92179@student.squ.edu.om, s91197@student.squ.edu.om, shaikh@squ.edu.om*

**Abstract –** The cybersecurity of modern systems has dramatically increased attention from both industrial and academia perspectives. In the recent era, the popularity of the blockchain-based system has traditionally been emergent among various industrial's sectors especially in supply chain management due to its streamlined nature. This reveals the importance of the quality aspects from a supply chain management perspective. Many industries realized the importance of having quality systems for supply chain management and logistics. The emergence of blockchain technology has created several potential innovations in handling and tracking business activities over the supply chain processes as specific. This paper shed the light on the blockchain and specifically on a smart contract technology which been used to handle the process of creation, verification and checking data over the supply chain management process. Then, touch upon the area of blockchain cybersecurity in the supply chain context. More and more, since the smart contract handles the transfer of data over different locations, then the security protection should be strong enough to secure the data and the assets from any attacks. Finally, the paper examines the main security attacks that affect the data on the blockchain and propose a solution.

* Corresponding Author : A. K. Shaikh, shaikh@squ.edu.om

# Introduction

To manage the total flow of a distribution channel from supplier to the ultimate user is known as supply chain management. Although most of the new supply chain management systems are complex and expensive, it considers one of the most important elements for many different disciplines of the economy. It works to handle the growing capacity and enhance the effectiveness of the product's flow [1]. A successful global supply chain management hinges upon the cohesive and harmonized management of four main flows, specifically, the product's flow, the processes, information and the cash. Moreover, it is used to stop any possible threats and ensure industry alignment. Today, technological innovations have made it easier and quicker to move products or full cargo from one place to another across the globe. Recently, a large number of companies manage and monitor their own IT systems for the supply chain processes. Yet, still difficult for those systems to collaboratively works to provide a clear vision of the scattered supply chain systems. Employing decentralized ledgers such as blockchain technology to build and sustain supply chain management will be a promising solution [2]. All the blockchain transactions are processed in a digital approach where the data will be stored, checked and proved online via the network without any approval from a central authority. All the data in the blockchain are stored in distributed nodes [3] and has some pros and cons in terms of security and transparency. This paper investigates how the blockchain technology helps to support the supply chain management processes [4], enhancing the process and data security and ensuring cost-effectiveness. The main contribution of this paper includes:

- Investigate the current challenges of blockchain-based supply chain and find out the ways to overcome them.
- Confer security concerns when utilizing blockchain technology in supply chain processes.
- Provide recommendations on common attacks against blockchain technology and its consequences on supply chain processes.

The rest of the paper is organized as below:

Section 2 presents the earlier research related to blockchain technology and supply chain management. Section 3 examines the pros and cons of blockchain technology when this technology is used in supply chain processes. Section 4 discusses the security attacks in the blockchain network and their implications on supply chain management processes and Section 5 concludes the research and presents the future research directions components, incorporating the applicable criteria that follow:

# Literature Review

### A. Blockchain Technology

Blockchain is a peer-to-peer cryptographic-based distributed ledger [5] that allows the creation of secure transactions permanently without third-party control [6]. Each block has to be validated by at least 51% of the nodes in the network [7]. There are many mechanisms of unanimity but the main four are: Proof of Work "POW", Proof of Stake "PoS", Practical Byzantine Fault Tolerance "PBFT", and Delegated Proof of Stake "DPoS". The most commonly used is PoW in Bitcoin and Ethereum platforms. PoW technique uses a solution to a puzzle to prove the trustiness of the data [8]. The process of solving the puzzle and processing must be done by checking all monetary transactions, which is called "mining"[9]. Once the data is proofed, only then the block can be created and have a hash value that links it to the previous block; Hence, a block is added to the blockchain [7]. The block then will be broadcasted to the nodes within the network [7]. Even though the data is publicly available for everyone who can access the blockchain, however, the data is encrypted and can only be decrypted by the intended parties involved. The cryptography of blockchain-based on asymmetric keys. Meaning, that the key used to encrypt the data is different from the key used to decrypt it. However, the two keys are mathematically linked and created the public and private keys. The public key is used to encrypt and validate the overall transactions, whereas the intended recipient within the network will use their private keys to decrypt the data [10]. Thus, the blocks created in the chain are secure from a confidentiality point of view, immutable and cannot be deleted or altered and its integrity is intact.

Blockchain mainly focused on Bitcoin and decentralized exchange of payment and cryptocurrencies. Then it evolved into Blockchain 2.0 in 2013 and focuses on decentralizing the whole market and exchanging the data and assets through smart contracts. Most of the smart contracts used today is running on Ethereum – an open decentralized platform [11] [12]. A smart contract is a digital contract that enables the trigger of an action based on predefined conditions to execute business transactions in a secure way [7] [13]. Accordingly, each created smart contract will be copied to each node in the blockchain in order to prevent tempering the contracts [11].

### B. Supply Chain Management

Nowadays, people thinking in supply chain sector been changed due to the changes that happened in the global industries. Currently, the industries can't compete productively while all supply chain parties are scattered across the country or even outside. The trend of having proper supply chain systems in place has grown dramatically over the few past decades. The historical initiatives of supply chain management can be traced to the beginnings of the apparel industry where the rapid response approach and then to the effective customers response of the grocery sector [14].

Various supply chain definitions have been presented in the past few years as the companies saw the concept gain popularity. According to Gao Z and his research team, Supply Chain management is a cooperation approach of business activities from the end consumers to the original goods or products providers [1]. Also, the APICS Dictionary defines it as "the full procedure from producing the raw resources up to the final usage of the final products [14].

A typical supply chain management processes consist of ordering and receiving the raw materials or products, supporting the customers' services and performance measurements. The management of multiple functions and processes across the enterprise is required to offer quick and quality actions to the supply chain events [1]. Logistic services normally play an integral role in delivering the intended values to consumers. The main intention of the logistic within the supply chain management is to get the products in the right and good conditions, at an appropriate time and with the lowest possible costs. Performance of the SCM is often measured based on value, dependability, rapidity, flexibility and price. Supply chain management is a complex process by nature and one of its major objectives of it is to reduce the risk associated with the process. The source of the risks within the SCM has been categorized into two types, specifically, holistic and atomistic. The holistic source of risk needs an overall supply chain process analysis to measure the risk associated. This type is highly preferred to the components which are complex, rare and of high value. Whereas, the atomistic focus is on selecting a specific part of the supply chain process to measure the risk associated. It is highly preferred to the components which are with low complexity, available and low value [15]. Although a large number of companies, accept the idea of properly managing the supply chain, the growth of this new sector is still low. The main factors behind this are mainly the lack of guidelines for establishing proper agreements with the supply chain parties and the failure to build proper procedures for assessing the agreements. As well as, a lack of trust within and outside the company. Lack of cohesive information systems and electronic business linking companies [14]. Across many activities that are possible to be tackled by the blockchain supply chain management also deserve special attention [16].

### C. Blockchain-based Supply Chain Management

Utilizing blockchain in a conventional supply chain process can overcome lots of the current operational obstacles and can provide several benefits as well. To start with, the traditional supply chain lacks the visibility between different parties as each one of them has its centralized database to manage their operation. Therefore, it raises other issues such as trust and security, as there is no transparency of the operations along the supply chain. This will increase the lateness of material or information flow between the different. Moreover, the current system is poor to prove its compliance with laws and regulations as the data can easily tamper. This may lead to significant health, safety, and environmental damages if the final produced product is not up to the standard [3].

The adoption of blockchain technology will resolve the visibility issue as the transactions will be publicly available for the concerned stakeholders. In addition, the blockchain will ensure the trust between the parties as the system is self-contained and not under the control of any third party. The exchange of data is secure using the public-key-infrastructure where the blocks of data are immutable. Thus, secure from attacker manipulation and ensure compliance with the laws and regulations. The research study [17] suggests the solution to use blockchain technology in the supply chain to improve security and integrity of chain transparency. Therefore, implementing blockchain technology in the supply chain context will make its processes faster, solid, intact, secure and more reliable [10] [18]. A business model approach is proposed by [19] for designing a blockchain-enabled supply chain. The model discussed the potential use cases and critical issues such as value proposition in terms of data integrity, security, ownership and privacy. A considerable amount of literature reviews distributed nodes and discusses the importance of utilizing blockchain technology to facilitate and robust the supply chain management process in several sectors, such as food, medical, and education. In addition, few studies examine the security perspective of the blockchain as a technology. However, the research study discusses the details the security aspect of binding blockchain in the supply chain context.

### D. *Conventional supply chain v/s Blockchain-based supply chain management*

The implementation of the supply chain management process starts by identifying the key players of the process [20]. As the objectives of this process are to create the most value for the entire supply chain network [21]. SCM is an integral part of other business' logistic processes from the end consumers to the suppliers that provide services, goods or information. The typical process of SCM consists of ordering or receiving the raw materials/products, supporting consumer services and performance measurement from the suppliers [1]. On the other hand, the cash flow process will be via customers to the suppliers for the delivery of the product. Moreover, the information flow will be through both suppliers and customers on a frequent based. Proper information flow also essential as both parties will be aware of how good or bad the relationship among them. Quick and quality responses to the supply chain management events are surely required to coordinate several functions across any enterprise. It illustrates the role of the supply chain in the business process. It provides the essentials for various functions and is responsible to manage the information flow [1]. A continuous effort from the company is essential, as the supply chain process integration is a dynamic process

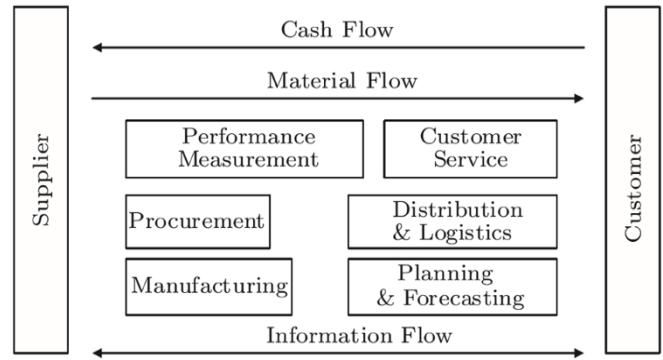

**Fig. 1** Illustrate the role of Supply chain management [1]

Adapting blockchain in supply chain processes can be done through the help of smart contracts. All supply chain parties should be the first part of the blockchain network, such as the supplier, producer, distributor, warehouse, and retailer. Each party will contribute to adding a new block to the blockchain whenever they perform their task. Each block is linked to the previous block coming from the last responsible party in the chain, and hence, creating the blockchain. Therefore, the data is cumulated and can be easily retrieved and traced.

The process starts when a supplier will provide the raw material to the producer as per their request. A new block will be added to the blockchain that has information about the raw materials such as the certificate of origin, batch data, order number, shipment date, and a barcode. Once the producer receives the raw material, this will
trigger the smart contract to check the shipment against the specifications, quantity, and quality mentioned in the order. If the raw materials received are as per the order, then the payment will be triggered by the smart contract to be processed to the supplier. Next, the producer will use the raw materials to produce products. Once the products are ready to be shipped, a new block will be added to the blockchain that has the previous product information provided by the supplier and added to it the production number. The newly added block will be linked to the previous block in the chain by the hash value. The products now are ready to be shipped to the warehouses. When the products are shipped, a new block will be added to the blockchain that has the shipment number in addition to the previous information. Once the warehouse receives the shipment, another smart contract will be triggered and match the supplier, order, invoice, and shipment and check the quality and quantity. If passed, then the inventory level will be updated in the warehouse, and the smart contract will trigger to process the of payment to the manufacturer. The warehouse then will ship the required products to the retailer as per their request and add a new block to the blockchain that has product received data, shipment date, and packaging barcode in addition to the previous information. The new block will be linked to previous blocks by a hash value. When the shipment is received by the retailer, another smart contract will be triggered to check if the received shipment is as per the order. If yes, a smart contract will be triggered to process the payment and the inventory levels will be updated. Then the retailer will add a new block to the blockchain that has the receive date, order number, and customer ID, which will be linked to previous blocks by a hash value. The end customer then can easily scan the barcode in the product to trace it back to its origins

## Blockchain's Security in Supply Chain Context
This section examines the pros and cons of blockchain technology when blockchain used within supply chain processes.

### A. Tamper-Proof Data
In the context of the supply chain, and because different stakeholders are involved along the process, it is essential to have a timestamp for each and every block added to the blockchain. The timestamp ensures the data integrity and no further modifications are allowed. Traditionally, the time stamp technique is performed on the blocks by a trusted server using the user's private key. However, if the server is compromised, this may result in signing and posting previous transactions. Therefore, the issue can be overcome by decentralizing and distributing the timestamping [22] [23].

### B. Single Point of Failure
One of the core features of blockchain technology is decentralization. The data is stored across all the nodes within the network. Each node will have a copy of the entire blockchain; which will prevent the one-point-failure issue. In addition, even if one or more of the nodes is down for several reasons such as maintenance, upgrade, or Denial of Service (DoS) attack, the data will still be available within the remaining nodes in the network. Thus, the blockchain ensures data availability. And as Ethereum is an open-source platform, its code is less prone to malicious changes as the code is under the supervision of many different parties. But like any other code, it may have some bugs or vulnerabilities due to human errors. All in all, insuring the data availability will boost the efficiency of the supply chain processes such as ordering the material in the right time and quantity, shipping the products when it is ready, and optimizing the stock levels in the warehouses [22]

### C. Privacy
Data privacy is essential in supply chain management as each party wants to protect its own data from being exposed to other stakeholders in the chain and maintain its competitive advantage [24]. Blockchain ensures the privacy of the data by using public-private keys (Asymmetric encryption). The keys are used to create a digital identity for each stakeholder to control the real-time data access [24] The mechanisms of exchanging the keys within the blockchain can be done by using one of two suites for Transport Layer Security (TLS), which are Rivest-Shamir-Adleman (RSA) or Elliptic Curve Diffie-Hellman Exchange (ECDHE). However, both of them were discouraged by the National Security Agency (NSA) in 2015. Some studies claim that this is due to the quick transition to quantum and post-quantum computing. In addition, the cryptographic primitives in blockchain are classified into two main categories: Optional and primary. The Optional category is used to enhance the anonymity and the privacy of blockchain-based transactions. Whereas the primary category contains the cryptographic hashes and standard digital signatures which are important for ensuring the blockchain as a global ledger with public verifiability, achievable consensus and tamper-proof [25]. Data privacy is fostered by the hash function that is used to sign the transactions in the blockchain. The level of data security depends on the hash functions used, such as Scrypt, SHA-256, and SHA256D. However, the hash function should be selected carefully based on the data sensitivity; as stronger the hash function may consume more time to process the transaction signing, which may delay the overall supply chain processes [22].

### D. Identity Management

Defining digital identities for supply chain stakeholder ensure data confidentiality, as the data can only be seen by authorized people. Thus, there are several ways to manage digital identities. For instance, a research study [26] proposed the model via a Secure Cryptography-based Clustering Mechanism where signature generation and signature verification processes are applied. To start with, a single entity can be responsible to manage and issue the digital identities for the stakeholder (centralized approach). However, if the supply chain stakeholders are located in different countries, each country will be responsible to generate a digital identity for their local companies. On the other hand, identity can be controlled by the users themselves (user-centric approach) [22].

### E. Smart Contract Management

While combining blockchain with the supply chain, smart contract is a key to facilitating the overall supply chain process. It is a piece of code that is triggered based on a specific pre-defined condition. As per the proposed system in this paper, the trigger of a smart contract will result in proceeding to the next process and will trigger the money payment process. All smart contracts are deployed and saved in the blockchain to prevent unauthorized amendments. However, the security of the smart contract depends on how robot the code is, as weak contracts are vulnerable and can be exploited by hackers. Therefore, they have to be written carefully. The triggers of smart contracts can be done manually or by pushed by oracles from an existing technology with the respective stakeholder to the smart contracts. Examples of such technology can be a company website, RFID, and IoT [22].

## Possible Security Attacks in Blockchain-based Supply Chain Context

This section discusses the most common security attacks in Peer-to-Peer networks (blockchain) and their implications on supply chain management processes. These attacks aim to disrupt the communication between the involved nodes within the network [27] and compromise data confidentiality, integrity, and availability. Therefore, when adopting blockchain technology in supply chain processes, these attacks should be considered, and build the network securely to avoid or minimize their effect.

### A. Sybil attacks

Sybil attack is considered one of the most famous and serious attacks in peer-to-peer networks like blockchain. The attack is performed when a user can control a large number of fake IDs within the network [28] which will lead to providing forged information to other legitimate nodes [27]. As a result, can compromise data confidentiality and integrity. The attack depends heavily on the vulnerabilities on the blockchain network, such as how cheap is the identity management defined, the level of trust linking the blocks, and how secure are the private keys. Thus, this attack is hard to prevent, but its effect can be minimized if proper security techniques are implemented.

### B. Eclipse attack

In this type of attack, an entity uses multiple identities to conspire and to cut off traffic passing through legitimate nodes as the main goal of this attack is to hide and eclipse the legitimate nodes from the network. This attack aims at the victim of public IP addresses. One of the main issues of the peer – 2 – peer networks is security and that's due to their distributed nature. This P2P network is subject to more complex attacks than the client-server networks. in this Eclipse attack, the attackers have control over a large part of the neighbors of good nodes in the network. Hence, they may manipulate the overlay network and control the majority of the nodes [27]. The attackers are able to control all the legitimate users' transactions and operations and hence, isolate them from the other connected nodes in the network. Moreover, they are also able to sort the user's view of the peer-to-peer network and force him to waste the power in unused directions in the blockchain networks.

As stated in the research study [29], this kind of attack can cause Denial of Service. This attack is hard to be eliminated but its effect of it can be minimized by applying the proper protections in place as limiting the access to the blockchain nodes to the highly authorized people. Eclipse attacks compromise the data availability, confidentiality and integrity hence, the effect of this attack on the supply chain management is very high as people will not be able to find the data or check the originality of the data or the resources then the financial loss will be high.

### C. 51% Attack

51% attack or Majority attack is one of the important cyber-attacks within the blockchain. Its mechanism is based on the ability of a hacker to get control of the majority – or more than half of the nodes within the network in order to make a false agreement in the voting system. The hacker cannot modify the existing data in the blockchain as he has no access to the participants' private keys. However, the attacker may create new fault blocks within the chain [30]. This attack is hardly performed in a public blockchain as the number of nodes within the network is huge. Thus, private networks are more vulnerable to this attack [29]. As a result, the PoW consumption power and electricity of the nodes which are under the control of the hacker is considered a cost for the network participants. The longer the attack will last, the more costly it is [30]. This attack compromises data confidentiality, integrity, and availability. Thus, it may incur financial losses to the stakeholders within the supply chain. In addition, the information flow will not be accessible, and therefore, it is hard to complete business transactions such as procurement, shipment, and performing quality check processes.

Based on the analysis of blockchain technology and security concerns, the paper suggests the use of Blockchain technology in supply chain management for a sustainable and secure process that will help to optimize the overall effectiveness of supply management processes. The use of blockchain would provide a better solution to the challenges faced by traditional supply chain management systems. In addition, the supply chain process can utilize the capabilities of the smart contract of Ethereum platform to monitor and manage the supply chain processes by the cryptography processes for better security protection of supply chain processes and transactions.

## Conclusion

This paper identified the challenges and their consequences in the traditional supply chain process and discussed how the blockchain-based supply chain can optimize the operations of supply chain processes throughout the value chain, and described how smart contracts in the Ethereum platform can help in end-to-end supply chain processes. In addition, the paper explored the blockchain's cybersecurity benefits and concerns when it is adopted in supply chain management. Finally. the paper examined the main security attacks on blockchain and its effects on supply chain processes and suggest the solution for sustainable and secure supply chain processes. For future research, additional security and interoperability issues need to be considered as currently, organizations are considering cloud-based blockchain technology in the supply chain management process.

## Reference

1. Gao, Z., et al., *Coc: A unified distributed ledger based supply chain management system.* Journal of Computer Science and Technology, 2018. **33**(2): p. 237-248.
2. Chang, Y., E. Iakovou, and W. Shi, *Blockchain in global supply chains and cross border trade: a critical synthesis of the state-of-the-art, challenges and opportunities.* International Journal of Production Research, 2020. **58**(7): p. 2082-2099.
3. Aich, S., et al. *A review on benefits of IoT integrated Blockchain based supply chain management implementations across different sectors with case study*. in *2019 21st international conference on advanced communication technology (ICACT)*. 2019. IEEE.
4. Korpela, K., J. Hallikas, and T. Dahlberg. *Digital supply chain transformation toward blockchain integration*. in *proceedings of the 50th Hawaii international conference on system sciences*. 2017.
5. Taylor, P.J., et al., *A systematic literature review of blockchain cyber security.* Digital Communications and Networks, 2020. **6**(2): p. 147-156.
6. Wang, Y., et al., *Making sense of blockchain technology: How will it transform supply chains?* International Journal of Production Economics, 2019. **211**: p. 221-236.
7. Lu, Q. and X. Xu, *Adaptable blockchain-based systems: A case study for product traceability.* IEEE Software, 2017. **34**(6): p. 21-27.
8. Li, X., et al., *A survey on the security of blockchain systems.* Future Generation Computer Systems, 2020. **107**: p. 841-853.
9. Holotescu, C. *Understanding blockchain opportunities and challenges*. in *Conference proceedings of» eLearning and Software for Education «(eLSE)*. 2018. " Carol I" National Defence University Publishing House.
10. Apte, S. and N. Petrovsky, *Will blockchain technology revolutionize excipient supply chain management?* Journal of Excipients and Food Chemicals, 2016. **7**(3): p. 910.
11. Cheng, J.-C., et al. *Blockchain and smart contract for digital certificate*. in *2018 IEEE international conference on applied system invention (ICASI)*. 2018. IEEE.
12. Shaikh, A.K., S.M. Alhashmi, and R. Parthiban. *A Semantic Decentralized Chord-Based Resource Discovery Model for Grid Computing*. in *Parallel and Distributed Systems (ICPADS 2011), IEEE 17th International Conference on*. 2011.

13. Brilliantova, V. and T.W. Thurner, *Blockchain and the future of energy.* Technology in Society, 2019. **57**: p. 38-45.
14. Lummus, R.R. and R.J. Vokurka, *Defining supply chain management: a historical perspective and practical guidelines.* Industrial management & data systems, 1999.
15. Kshetri, N., *1 Blockchain's roles in meeting key supply chain management objectives.* International Journal of Information Management, 2018. **39**: p. 80-89.
16. Nakasumi, M. *Information sharing for supply chain management based on block chain technology*. in *2017 IEEE 19th conference on business informatics (CBI)*. 2017. IEEE.
17. Saberi, S., et al., *Blockchain technology and its relationships to sustainable supply chain management.* International Journal of Production Research, 2019. **57**(7): p. 2117-2135.
18. Xu, L., et al. *Binding the physical and cyber worlds: A blockchain approach for cargo supply chain security enhancement*. in *2018 IEEE International Symposium on Technologies for Homeland Security (HST)*. 2018. IEEE.
19. Wang, Y., C.H. Chen, and A. Zghari-Sales, *Designing a blockchain enabled supply chain.* International Journal of Production Research, 2020: p. 1-26.
20. Chen, H., P.J. Daugherty, and T.D. Landry, *Supply chain process integration: a theoretical framework.* Journal of business logistics, 2009. **30**(2): p. 27-46.
21. Croxton, K.L., et al., *The supply chain management processes.* The International Journal of Logistics Management, 2001. **12**(2): p. 13-36.
22. Fraga-Lamas, P. and T.M. Fernández-Caramés, *A review on blockchain technologies for an advanced and cyber-resilient automotive industry.* IEEE Access, 2019. **7**: p. 17578-17598.
23. Shaikh, A., S. Alhashmi, and R. Parthiban, *A Semantic Impact in Decentralized Resource Discovery Mechanism for Grid Computing Environments*, in *Algorithms and Architectures for Parallel Processing*. 2012, Springer Berlin Heidelberg. p. 206-216.
24. Chen, S., et al. *A blockchain-based supply chain quality management framework*. in *2017 IEEE 14th International Conference on e-Business Engineering (ICEBE)*. 2017. IEEE.
25. Wang, L., et al., *Cryptographic primitives in blockchains.* Journal of Network and Computer Applications, 2019. **127**: p. 43-58.
26. MOHINDRA, A.R. and C. GANDHI, *A secure cryptography based clustering mechanism for improving the data transmission in MANET.* Walailak Journal of Science and Technology (WJST), 2021. **18**(6): p. 8987 (18 pages)-8987 (18 pages).
27. de Asís López-Fuentes, F., I. Eugui-De-Alba, and O.M. Ortíz-Ruiz, *Evaluating P2P networks against eclipse attacks.* Procedia Technology, 2012. **3**: p. 61-68.
28. Shareh, M.B., et al., *Preventing Sybil attacks in P2P file sharing networks based on the evolutionary game model.* Information Sciences, 2019. **470**: p. 94-108.
29. Liang, G., et al., *Distributed blockchain-based data protection framework for modern power systems against cyber attacks.* IEEE Transactions on Smart Grid, 2018. **10**(3): p. 3162-3173.
30. Guégan, D. and C. Hénot, *A probative value for authentication use case blockchain.* Digital Finance, 2019. **1**(1-4): p. 91-115.

# Biography

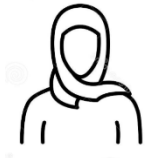

**Ms Al-Lawati** is a final year MSc student in the Department of Information Systems, College of Economics & Political Science at Sultan Qaboos University Her master research investigates Factors Affecting Undergraduate Students Academic Performance Resulting in Falling Under Probation, Her research interest include Blockchain, Supply chain, Peer to Peer network
.

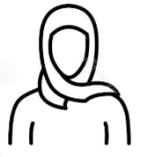

**Ms Al Busaid** Lawati is a final year MSc student in the Department of Information Systems, College of Economics & Political Science at Sultan Qaboos University Her master research investigates Fostering knowledge management for institutional development. Her research interest include Blockchain, Supply chain and Smart contract,
.

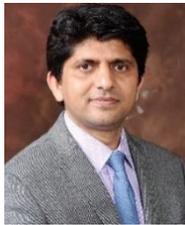

**Abdul Khalique Shaikh is** an Assistant Professor associated with the Department of Information Systems at Sultan Qaboos University Muscat Oman. He has received his PhD degree from a highly reputable Australian University Monash in 2013 and Master degree from University of Detroit USA with distinction. He has a strong skill set and experience gained from working in both academia and industry that endow him with a broad background and a unique perspective on research & teaching. His research interest includes Social Network Analytics, Big Data Analytics, Data Science, Data Governance, E-participation and Blockchain Technology.